\RequirePackage{fixltx2e}
\documentclass[manuscript=article]{achemso}
\AtBeginDocument{\let\latin\relax}
\usepackage[T1]{fontenc}       
\usepackage{lmodern}
\usepackage{chemfig}
\usepackage{lineno}
\usepackage{amsmath}
\usepackage{bm}
\usepackage{color}
\usepackage{graphicx}
\usepackage{subfig}
\usepackage{chemmacros}
\usepackage{pgfplots}
\usepackage{tikz}
\usepackage{units}
\usepackage{multirow}
\usepackage{booktabs}
\usepackage{natbib}
\usepackage{soul}
\usepackage{xr-hyper}
\usepackage{hyperref}
\usepackage{epstopdf}
\usepackage{listings}
\usepackage{caption}
\usepackage{amsfonts}
\usepackage{dsfont}
\pgfplotsset{compat=1.8}
\SectionNumbersOn

\tikzset{
  elmove/.style={-{Stealth[#1]},shorten >=3pt,shorten <=2pt}
}

\def\affilStuttgart{\affiliation{Max-Planck-Institut f{\"u}r Festk{\"o}rperforschung, Heisenbergstra{\ss}e 1, 70569 Stuttgart, Germany}}

\author{Giovanni Li Manni}
\affilStuttgart
\email{g.limanni@fkf.mpg.de}

\author{Ali Alavi}
\affilStuttgart
\email{a.alavi@fkf.mpg.de}

\title{Understanding the mechanism stabilizing intermediate spin states in Fe(II)-Porphyrin}
\abbreviations{}
\keywords{}
\date{\today}

\begin{document}

\begin{abstract}
	Spin fluctuations in Fe(II)-porphyrins are at the heart of heme-proteins functionality.
        Despite significant progress in porphyrin chemistry, the mechanisms that rule
	spin state stabilization remain elusive.
	Here, it is demonstrated by using multiconfigurational quantum chemical approaches,
	including the novel Stochastic-CASSCF method, that
	electron delocalization between the metal centre and the $\pi$ system of the
	macrocycle differentially stabilizes the triplet spin states over the quintet.
	This delocalization takes place {\em via} charge-transfer excitations,
	involving the out-of-plane iron $d$ orbitals,
	key linking orbitals between metal and macrocycle.
	Through a correlated \textit{breathing mechanism} the 3d electrons can make transitions towards the $\pi$ orbitals of the macrocycle.
	This guarantees a strong coupling between the on-site radial correlation
        on the metal and electron delocalization.
	Opposite-spin 3d electrons of the triplet can effectively reduce electron repulsion in this manner.
	Constraining the out-of-plane orbitals from breathing hinders delocalization and reverses the spin ordering.
        Our results find a qualitative analogue in Kekul{\'e} resonance structures involving also the metal centre.
\end{abstract}
	\newpage

\section{Introduction }
	\label{Introduction}
	Metal-porphyrins are versatile chemical species which biological systems make abundant use of,
	with Mg(II)-porphyrins and Fe(II)-porphyrins being the most common in nature. The latter are used in a 
        number of vital functions in aerobic life, including dioxygen transport and reduction.
	From an electronic point of view Mg(II) porphyrins are closed shell
	diamagnetic compounds, while Fe(II)-porphyrins with a $d^{6}$ configuration at the metal centre 
	may show a multitude of low-lying electronic states. 
	The high-spin (quintet), the intermediate-spin (triplet) and the low-spin (singlet) states are near degenerate and,
	depending on the coordinating ligands, geometry and thermodynamical conditions, their relative energy may easily change.
	Spin changes are the key feature that enables enzymatic and biomimetic reactions involving Fe-porphyrins.
	Molecular and electron transport as well as metabolic reactions take place thanks to the facile spin
	and oxidation state changes of these compounds.
	The oxidative oxygenation (insertion of one oxygen in a C--H bond) by the cytochrome P450s
	is an example\cite{Sono1996,Shaik2004,Shaik2002}.
	In this reaction, the Fe(II)-porphyrin represents the active species that binds molecular
	oxygen and weakens its bond, forming the actual oxo-species that proceeds to the oxygenation.
	It is then crucial from a mechanistic point of view to understand the electronic structure 
	of Fe(II)-porphyrins and the main elements that stabilize one spin state over the others.

	The first \textit{ab initio} calculations on the free-base porphyrin, that can be regarded
	as the parent compound for such systems, were done by Alml{\"o}f\cite{Almlof1974}.
	Later numerous density functional theory calculations were carried out in order
	to explore the electronic properties of metal porphyrins\cite{Ziegler1993,Baerends1993}.
	In spite of a large amount of experimental and theoretical data,
	many questions are still unanswered regarding their electronic properties and reactivity.
	For instance, a definitive assignment of the ground state of four-coordinated ferrous porphyrins
	is still missing and the main ingredients governing ground state electron configuration unknown to date.
	A $^3A_{2g}$ ground state with configuration $(d_{x^{2}-y^{2}})^2(d_{xz},d_{yz})^2(d_{z^2})^2$ 
   	($d_{xy}$ refers to the anti-bonding orbital pointing at the N atoms)
	was suggested by M{\"o}ssbauer \cite{Collman1975,Collman1978}, magnetic\cite{Boyd1979} and H-NMR
	\cite{Goff1977,Mispelter1980} measurements of the Fe(II)-tetraphenilporphyrin (FeTPP). 
 	A $^3E_g$ state was suggested by Raman spectroscopy\cite{Kitagawa1979}, with
	configuration $(d_{x^{2}-y^{2}})^2(d_{xz},d_{yz})^3$ $(d_{z^2})^1$ for the Fe(II)-octaethylporphyrin (FeOEP). 
 	A high-spin state was reported for the octamethyl-tetrabenzporphyrin-iron(II)\cite{Sams1974}.
        Many factors, such as functionalization of the macrocycle and solvation, 
	may affect the relative ordering of the low-lying states. 

	A large number of theoretical studies are available for model systems of the ferrous porphyrin
        \cite{Matsuzawa1995,Kozlowski1998,Scheiner2002,Smith2005,Chen2011,Choe1999,Choe1998,Vancoillie2011,Pierloot2017,Phung2016,Radon2008,Vancoillie2010,Radon2010,Radon2014}.
	Density Functional Theory (DFT) predicts a triplet ground state for the Fe(II)-porphyrin,
	although still consensus has yet to be reached for the specific symmetry of the state.
        Swart \textit{et al.} found a high sensitivity of spin gaps on the type of functional used
	in density functional approximations\cite{Swart2004}.
        The OPBE functional\cite{Perdew1996,Cohen2001} predicted a $^{3}E_{g}$ ground state for the ferrous porphyrin,
	with the $^{3}A_{2g}$ and the $^{5}A_{1g}$ at 4.0 and 7.2 kcal/mol above respectively.
        Interestingly, other functionals (BP86 and B3LYP) predicted a $^{3}A_{2g}$ ground state with the
        $^{3}E_{g}$ at 1.8 and 6.2 kcal/mol above respectively\cite{Kozlowski1998}.

	A completely different scenario  is depicted by wave-function theory based approaches, including high-level methodologies generally regarded
as being more reliable than DFT.
	The Restricted Open-shell Hartree-Fock method, ROHF, predicts the high-spin $^{5}A_{1g}$ ground state.
	The gap between the quintet and the triplet spin states shrinks when post-SCF methods are used,
	however, the spin-ordering remains in favor of the high-spin state,
	suggesting a systematic error in the theoretical framework.
	Pierloot has extensively studied these systems by multiconfigurational methods and how spin gap predictions 
	depend on the choice of the active space\cite{Radon2010,Vancoillie2010,Radon2008,Pierloot2017,Phung2016}.
	A definitive argument on the mechanism stabilizing the triplet spin state was not suggested.
	Transition metal spin chemistry has always been challenging for
	quantum chemical methods\cite{Swart2008,Swart2013,Harvey2013}
	and a simple and reliable theoretical approach for spin-dependent properties is still not available\cite{Burke2017}.
	In addition to the lack of consensus among the theoretical methodologies, a more fundamental question is still
	unanswered, \textit{what factors dictate the relative stabilization of the competing spin states?}

\begin{figure}
    \centering%
    \includegraphics[width=10cm]{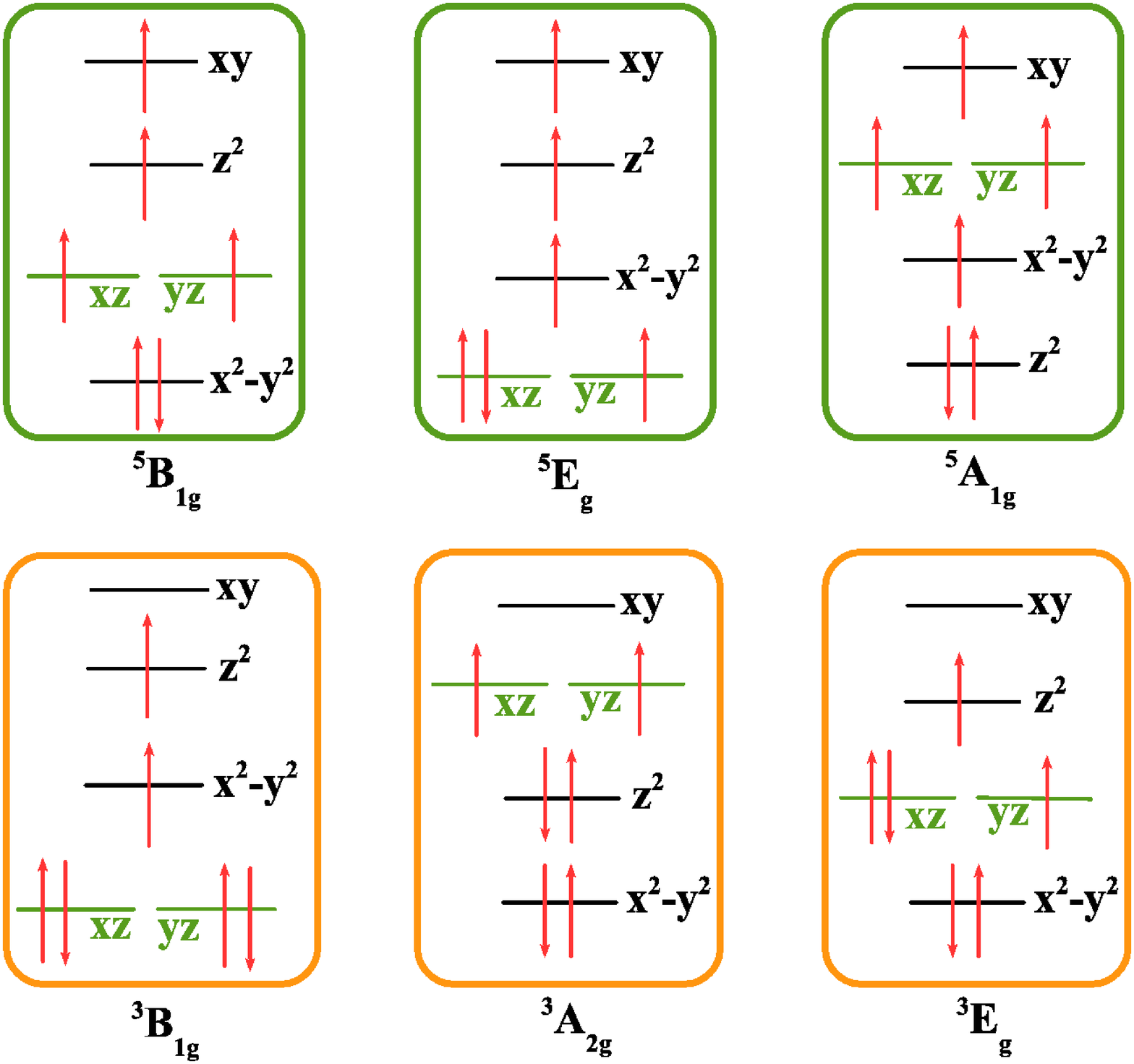}
    \caption{Dominant electron configuration of the six competing spin states of Fe(II)-porphyrin.}%
    \label{Fig:6states}
\end{figure}

	In this report we will show in great detail the mechanisms that rule spin gaps in the bare ferrous porphyrin, by analyzing
	the six low-lying states, $^{3}B_{1g}$, $^{3}A_{2g}$, $^{3}E_{g}$, $^{5}A_{1g}$, $^{5}B_{2g}$ and $^{5}E_{g}$, 
	of the Fe(II)-porphyrin (Figure~\ref{Fig:6states}).
	The Stochastic Complete Active Space Self-Consistent Field method, Stochastic-CASSCF
	\cite{limanni2016,Roos1980a,Roos1980b,Siegbahn1980,Siegbahn1981,Roos2007},
	is the method of choice for this investigation.
	The CASSCF represents a simple and natural way to systematically probe
	correlation channels in	correlated molecular systems.
        The core concept of CASSCF is the active space, a list of ``critical'' orbitals with their electrons for which a 
	complete many-body expansion is generated and orbitals are variationally optimized under 
	the field generated by the multiconfigurational wave-function, removing any bias related
	to the choice of the trial orbitals.
	Equipped with the Stochastic-CASSCF method we have been able to identify the main correlation effects that stabilize
	the intermediate spin-state over the quintet spin-state and have been able to establish that
	important communication pathways between the aromatic macrocycle and the metal centre
	exist only for the low-lying triplet states.
	Qualitatively these correlation channels can be described as Kekul{\'e}
	resonance structures involving also the metal centre.

\section{Results} \label{Results}
Energy splittings at the various levels of theory and VTZP basis set are summarized in Figure~\ref{Fig:VTZP} 
(additional results available in the Supplementary Material).
Energy gaps between the quintet spin states are rather insensitive to the choice of the active space,
basis set and dynamic correlation correction via second order perturbation theory,
CASPT2\cite{andersson1990b,andersson1992b,andersson1993,finley1998,Ghigo2004,Pierloot2003,Pierloot2008,Pierloot2006}.
When turning our attention to the triplet spin states the scenario is rather different. 
     \begin{figure}
         \centering%
         \includegraphics[width=16cm]{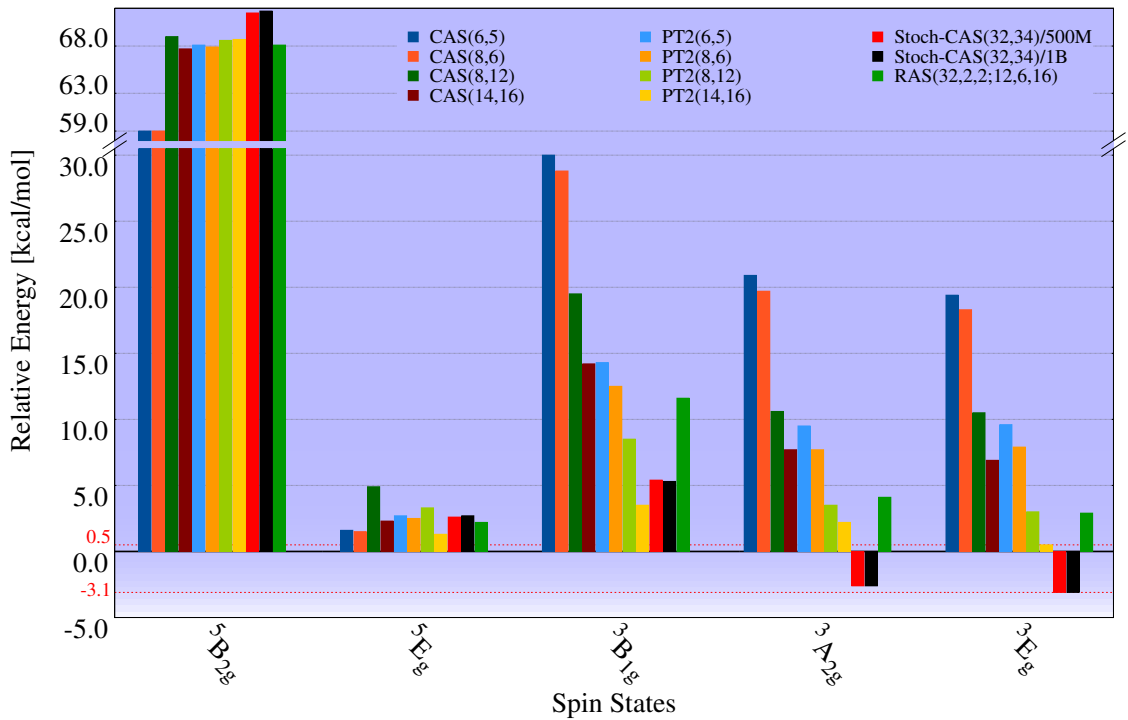}
	 \caption{Energy splittings relative to the $^{5}A_{1g}$ state in VTZP basis set. 
		  The upper red dashed line marks the lowest relative energy value that one 
		  could obtain with conventional CASPT2(14,16). The lower red dashed lines 
    		  marks the lowest relative energy value for the Stochastic-CASSCF(32,34) method.%
         }%
         \label{Fig:VTZP}
     \end{figure}
A strong dependence of the relative energy with respect to the active space and method of choice is observed.
CASPT2 spin gaps are affected both by the under-lying active space and basis set of choice.
Enlarging the active space reduces the gap between the triplet states and the $^{5}A_{1g}$ state.
Perturbative correction to the second order for any choice of active space approximately halves the triplet-quintet gap.
The smallest CASSCF(6,5) and CASSCF(8,6) place the lowest triplet, $^{3}E_{g}$, at $\sim18$~kcal/mol
above the $^{5}A_{1g}$ state.
When perturbative correction is added the $^{3}A_{2g}$ becomes the lowest
triplet state, lying at $\sim9$~kcal/mol above the quintet ground state.
The larger CASPT2(14,16) places the triplet $^{3}E_{g}$ at $0.5$ kcal/mol above the $^{5}A_{1g}$
ground state (3.2 kcal/mol with VDZP basis set).
These results clearly show that conventional CASPT2 is still not converged and the triplet might
(\textit{and will}) be further stabilized by more accurate methods, namely a larger under-lying active space.

The most notable result summarized in Figure~\ref{Fig:VTZP}
is the energy splitting obtained by the large CASSCF(32,34).
At this level of theory both the $^{3}E_{g}$ ($-3.1$~kcal/mol) and the $^{3}A_{2g}$ ($-2.6$~kcal/mol)
states are below the $^{5}A_{1g}$ state,
setting the $^{3}E_{g}$ as the ground state for this system. The $^{3}A_{2g}$ is only $0.5$~kcal/mol above the $^{3}E_{g}$.
The energy lowering obtained by this large active space is substantial.

Surprisingly, the restricted active space approach, RASSCF(32,34)
sets the triplet states again above the $^{5}A_{1g}$ state
(2.9 and 4.1~kcal/mol for the $^{3}E_{g}$ and the $^{3}A_{2g}$ state respectively).
Truncation of the excitation level to only single and double excitations from RAS1 and to RAS3
obviously has a strong unfavorable impact on the triplet-quintet splittings.
Within the Stochastic-CASSCF approach the impact of the target number of walkers is nearly negligible.
Variations of less than 0.1~kcal/mol were observed by enlarging the walker population from $500$ million to $1$ billion
(before convergence is reached, higher accuracy is expected for a larger walker population).
The large CASSCF calculations seem to be able to circumvent the limitations of smaller active spaces,
truncated CI expansions (RAS case) and perturbatively-corrected results.
Active space size limitations are substantial and RAS type of truncations do not represent a solution.
CASPT2 energy estimates with small active space reference wave functions
are not to be considered reliable for this class of compounds. 

    \begin{figure}
        \captionsetup[subfigure]{labelformat=empty,farskip=-25pt,captionskip=3pt}
        \centering%
         \includegraphics[width=16cm]{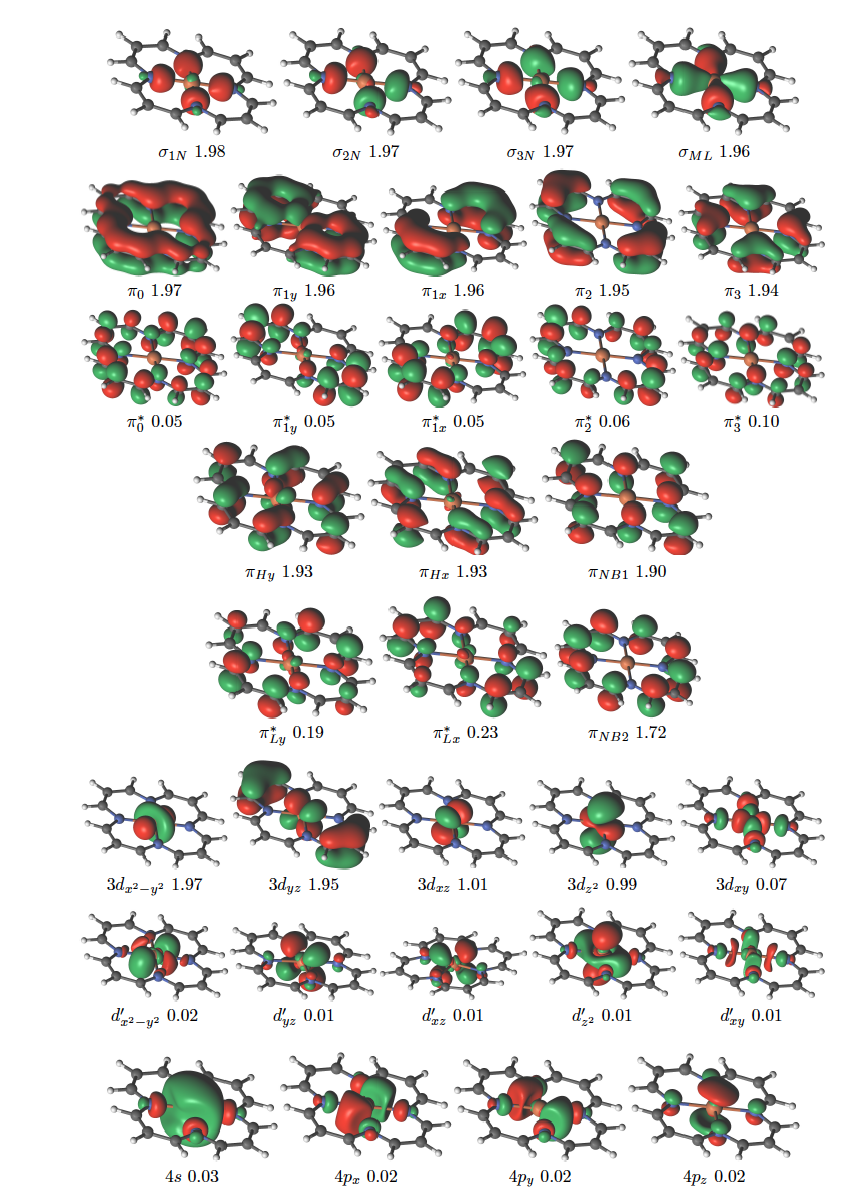}
        \caption{%
            Stochastic-CAS(32,34) active natural orbitals for the $^{3}E_{g}$ state of the Fe(II)-porphyrin model system
	    and their occupation numbers.
        }
        \label{Fig:3Eg_NatOrb3234}
    \end{figure}

From the inspection of the CASSCF(32,34) natural orbitals for the $^{3}E_{g}$ state 
a strong mixing of the $3d_{yz}$ orbital (and $3d_{xz}$ for the degenerate state)
with $\pi$ orbitals of the macrocycle are observed (see orbitals $\pi_{1y}$ and $\pi_{Hy}$ of
Figure~\ref{Fig:3Eg_NatOrb3234} as an example).
This mixing is missing in the $^{5}A_{1g}$ state (natural orbitals for the quintet spin
state are depicted in the Supplementary Material).
The mixing of the doubly occupied $3d_{yz}$ and the symmetry allowed $\pi$-orbitals for the triplet state is not fortuitous.
It means that off-diagonal elements in the one-body density matrix are quite large and the eigen-vectors 
leading to the natural orbitals will have large contributions from the metal centre and the macrocycle.
Large off-diagonal matrix elements can arise only when the wave function is multiconfigurational, 
with a strong entanglement between orbitals related to those large off-diagonal matrix elements and,
with cumulatively large amplitudes for \textit{charge-transfer} (CT) determinants
(CT determinants are those where starting from a given determinant
electrons are excited from the occupied $3d$ orbitals to the empty $\pi^{*}$ orbitals
and from the occupied $\pi$ orbitals to empty $3d$ orbitals).
Therefore, in the $^{3}E_{g}$ state, $\pi$ orbitals are strongly correlated to the $3d_{yz}$
orbital (and $3d_{xz}$) via CT determinants.
This property, that is not present in the $^{5}A_{1g}$ state, is the driving force that stabilizes
the triplet over the quintet spin state.

\begin{figure}
         \centering%
         \includegraphics[width=12cm]{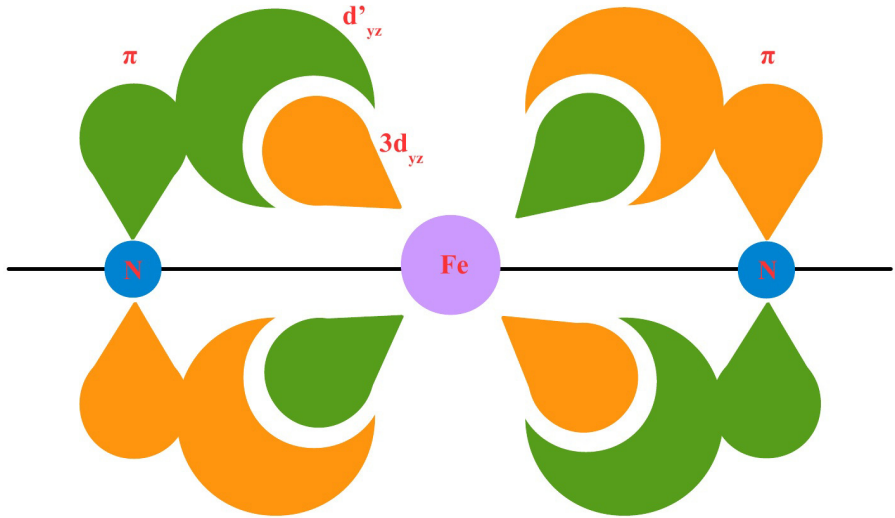}
	 \caption{The correlating $d'$ orbitals provide a correlation pathway for connecting $3d$
		  and $\pi$ orbitals in the iron complex.%
         }%
         \label{Fig:3d_bridging}
     \end{figure}

Adding the \textit{double-shell} $d'$ orbitals into the active space allows the triplet state
to differentially reduce on-site electron repulsion by exciting electrons out of the
doubly occupied $3d$ orbitals, into the $d'$ shell.
This effect is referred to as ``\textit{radial correlation}'' or ``\textit{breathing}''.
However, the $d'$ orbitals also open a ``correlation pathway'', a communication channel
between the $\pi$ orbitals and the 3d orbitals (see Figure~\ref{Fig:3d_bridging}).
Via the correlating $d'$ orbitals valence out-of-plane electrons are able to breathe out,
expand towards the $\pi$ orbitals of the macrocycle.
This guarantees a strong coupling between the on-site radial correlation on the metal centre
and electron delocalisation into the macrocycle.
Constraining the out-of-plane orbitals from breathing hinders delocalization and leads to the reversal of the spin ordering.
This confirms the crucial role of the \textit{breathing effect} in the spin chemistry of these systems.
In the CAS(14,16) four $\pi$ orbitals of the macrocycle are added into the active space,
$\pi_{Hx}$, $\pi_{Hy}$, $\pi_{Lx}^{*}$ and $\pi_{Ly}^{*}$
($x$ and $y$ subscripts refer to the symmetry of these orbitals) and are
explicitly correlated to the 3d and $d'$ orbitals at the metal centre.
They are the two degenerate HOMO ($H$ subscript) and two degenerate LUMO ($L$ subscript) orbitals of the free-base porphyrin.
Including these $\pi$ orbitals in the active space in absence of the double-shell
orbitals, leads to less populated anti-bonding $\pi$ orbitals. 
This difference is entirely due to the $d'$ double-shell. This is a very interesting finding.
Correlating $d'$ and $\pi$ orbitals synergistically favour electron delocalization and
reduce on-site electron repulsion at the metal centre.
This synergic effect arises from the coupling through the Hamiltonian operator of determinants of the type
$|3d\rightarrow d'\rangle$, $|\pi_{H} \rightarrow d'\rangle$, $|d'\rightarrow \pi_{L}^{*}\rangle$,
$|\pi_{H}\rightarrow \pi_{L}^{*}\rangle$ and $|\pi_{L}^{*}\rightarrow d'\rangle$.
This type of excitations are somehow included, although only up to the second order both in the RASSCF(32,34) and the CASPT2(8,12).
As neither CASPT2(8,12) or RAS(32,34) provided converged energetics,
we conclude that it is not sufficient to correlate these orbitals solely via singly- and doubly-substituted determinants.
Higher order excitations are responsible for the stabilization of the triplet over the quintet spin state.
Also, these excitations have been included in the CAS(14,16), but there full delocalization is not
explicitely accounted for in the method 
(most of the $\pi$ orbitals are not in the active space) and as a consequence the triplet lies still above the quintet spin state.
The important excitations, however, have been included in the large CASSCF(32,34) calculations.

\begin{figure}
     \centering%
     \includegraphics[width=16cm]{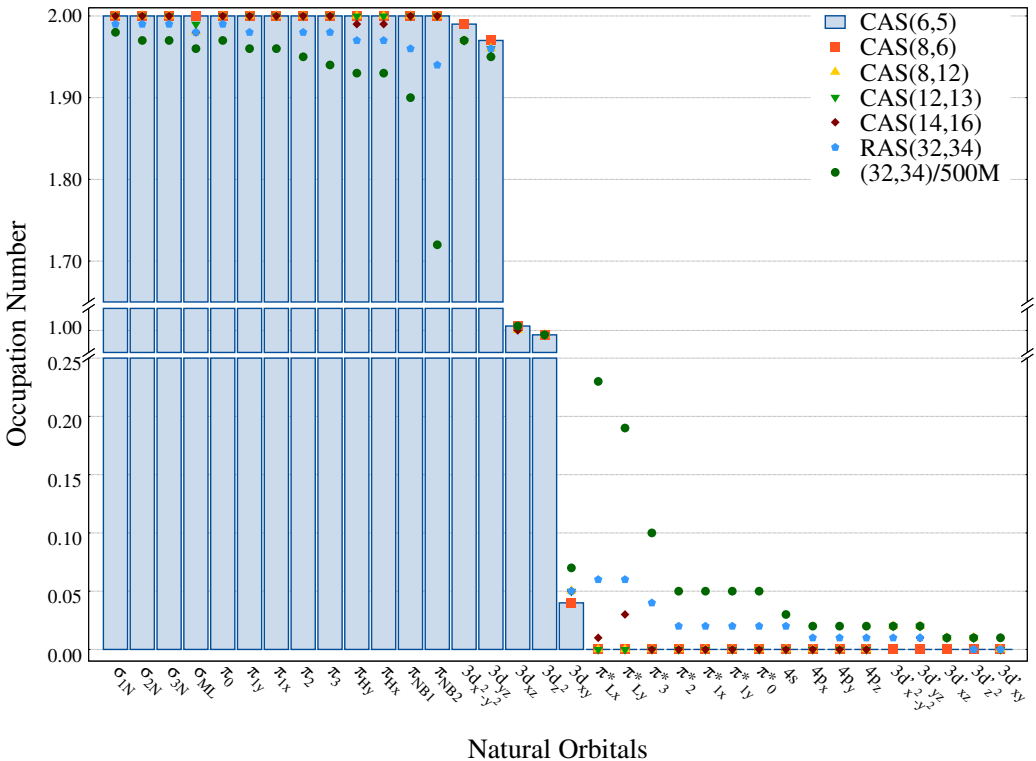}
     \caption{Natural orbitals occupation numbers for the $^{3}E_{g}$ state within the VTZP basis set for
	      different choices of active spaces. Bars are related to the minimum CAS(6,5) active space
	      that can be considered the reference for larger active spaces.%
     }%
     \label{Fig:NatOrbs_3Eg}
 \end{figure}

\begin{figure}
         \centering%
         \includegraphics[width=16cm]{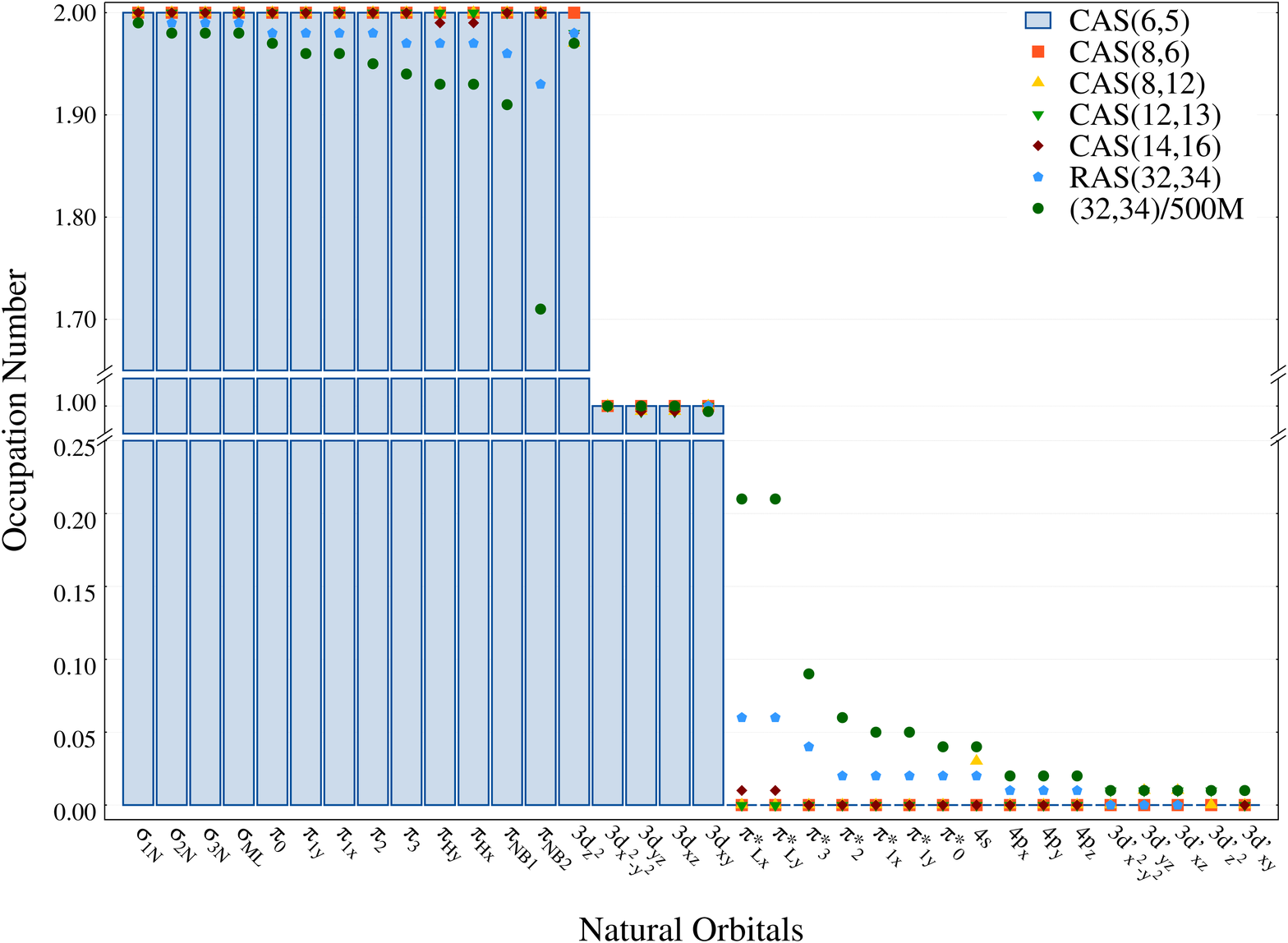}
	 \caption{Natural orbitals occupation numbers for the $^{5}A_{1g}$ state within the VTZP basis set for
		  different choices of active space. Bars are related to the minimum CAS(6,5).%
         }%
         \label{Fig:NatOrbs_5A1g}
     \end{figure}

Inclusion of the entire $\pi$-system in the active space and the complete CI expansion of the  CASSCF(32,34)
reveal another important result.
The $\pi$ non-bonding orbitals, $\pi_{NB1}$ and $\pi_{NB2}$, have rather low occupation numbers,
1.90 and 1.72 respectively, both for the $^{3}E_{g}$ and the $^{5}A_{1g}$ states.
At the same time $\pi_{Lx}^{*}$ and $\pi_{Ly}^{*}$ orbitals reach relatively high occupation numbers
(see Figure~\ref{Fig:NatOrbs_3Eg} and Figure~\ref{Fig:NatOrbs_5A1g}).

These results sensibly differ from the RAS(32,34) results and demonstrates that the
CAS(32,34) wave function is substantially different from the other wave functions here analyzed,
with some of the $\pi$ orbitals heavily involved in the correlation for this system.
The large CASSCF(32,34) provides a more accurate description of the electron delocalization
(conjugation and aromaticity) of the macrocycle. It also better accounts for CT determinants. 
As a consequence a more realistic field around the metal centre is created that, 
in response, differentially stabilizes the triplet over the quintet spin state.
This effect is achieved only when the wave function is relaxed with respect to CI and orbital parameters and, 
contains higher order excitations coupled via the true Hamiltonian operator.

    A deeper analysis of the large Stochastic-CAS(32,34) wave function corroborates the previous findings.
    Ligand-to-metal charge-transfer (LMCT) excitations, $\pi \rightarrow (3d_{xz},3d_{yz})$, contribute
    for $\sim 1\%$ for the $^{3}E_{g}$ state.
    Numerous LMCT excitations of the type $\pi \rightarrow (d'_{xz},d'_{yz})$,
    and metal-to-ligand charge-transfer (MLCT) excitations of the type $(d'_{xz},d'_{yz})\rightarrow \pi^{*}$ and
    $(3d_{xz},3d_{yz})$ $\rightarrow \pi^{*}$,
    contribute to the CAS(32,34) wave function of the $^{3}E_{g}$ state with weights around 0.1-0.5\%.
    CT determinants were already observed and reported in our previous work.
    However, in the present work $\pi \rightarrow (d'_{xz},d'_{yz})$ and $(d'_{xz},d'_{yz})\rightarrow \pi^{*}$ are also observed.
    This finding reinforces the above statement, double-shell $d'$ orbitals contribute to the ``radial correlation''
    and also serve as a communication channel bridging the gap between metal centre and macrocycle orbitals.
    They are actively involved in the delocalization of the valence electrons.
Valence out-of-plane 3d electrons expand with a \textit{breathing mechanism} via the correlating $d'$ orbitals,
which have a larger overlap with the $\pi$ orbitals of the macrocycle, and thus delocalize into the $\pi$ system.

    Charge-transfer excitations are rare for the quintet spin state.
    The LMCT $\pi\rightarrow 3d_{xz,yz}$ excitations contribute for only 0.3\% to the wave function. 
    MLCT excitations are even less representative of the wave function, 
    with the $3d_{xz,yz}\rightarrow \pi^{*}$ contributing for only 0.05\%.
    The $^{3}E_{g}$ state of the ferrous porphyrin is characterized by
    an important interaction between $\pi$ electrons and valence electrons at the metal centre.
    This feature is unique to porphyrinoids hosting transition metal centres.
    In fact, the Mg(II) porphyrin reported in our previous work did not show any interaction (via CT excitations)
    between the magnesium orbitals and the $\pi - \pi^*$ system of the aromatic macrocycle.
    In the Mg(II) compound doubly occupied orbitals of the magnesium lie low
    in the energy spectrum and the virtual manifold too high in energy to mix with the orbitals of the $\pi - \pi^*$ system.
    In the Fe(II)-compound 3d orbitals mix well both in terms of symmetry/overlap
    (considering the role of the $d'$ shell) and energy.

\section{Discussion} \label{Conclusions}
It is demonstrated by means of multiconfigurational approaches that for the 
bare Fe(II)-porphyrin the triplet spin state is stabilized over the quintet 
spin via metal-to-ligand and ligand-to-metal charge-transfer excitations. 
These are numerous in the intermediate spin state while being extremely rare in the quintet spin state.

Previous quantum chemical simulation of the same CASSCF type, but based on
smaller active spaces, do not describe adequately the above mentioned charge-transfer
configurations (they have too small an amplitude  or do not exist at all).
None of the smaller correlated methods here discussed shows $3d_{xz,yz}$/$\pi$ orbital mixing.
Simply stated previous methods do not show electron delocalization between metal centre and macrocycle and,
as a consequence the high spin states are overstabilized.
Previous multiconfigurational methods fail in capturing simultaneously ring correlation (aromaticity),
correlation at the metal centre (radial correlation) and the interaction between metal centre and macrocycle 
via charge-transfer excitations which differentially stabilise the intermediate spin states over the high spin ones. 

The interaction between metal and macrocycle is rather complicated.
In the large CAS(32,34) wave function, we observe non-negligible charge-transfer 
electron configurations coupling directly $\pi$ and $3d$ orbitals.
Higher order interactions between macrocycle and metal centre are also present via the double-shell $d'$ orbitals.
The double-shell orbitals play a dual role, they account for radial correlation at the metal centre and
build a bridging pathway between $\pi$ orbitals at the macrocycle and valence orbitals at the metal centre.
A \textit{breathing mechanism} can be invoked for rationalize the stabilization of the triplet spin state over the quintet.
Via the correlating $d'$ orbitals, which provide the necessary breathing mechanism, the valence electrons can more easily delocalize 
into the $\pi$ system, and to a far greater extent than the regular (non-breathing) 3d orbitals would allow.

\begin{figure}
         \centering%
         \includegraphics[width=16cm]{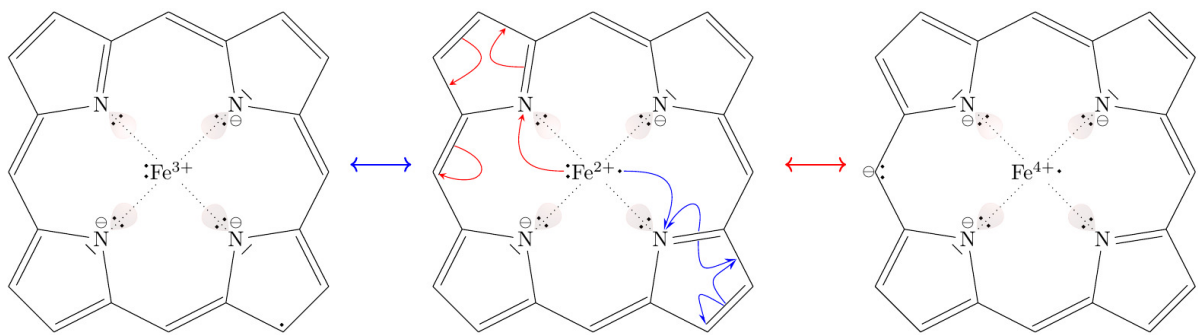}
	 \caption{Kekul{\'e} resonance structures involving movement of iron valence
		  electrons of the $3d_{xz}$ and $3d_{yz}$ orbitals.
         }%
         \label{Fig:mechanism}
     \end{figure}

The charge-transfer configurations can be qualitatively described as Kekul{\'e} aromatic resonance structures,
involving movement of the $\pi$ orbitals of the macrocycle as well as the valence electrons
on the $3d_{xz}$ and $3d_{yz}$ orbitals of the metal centre as described in Figure~\ref{Fig:mechanism}.
To the best of our knowledge this is the first quantum chemical investigation that explicitly shows how
the electron delocalization between metal centre and macrocycle
is the key feature for the stabilization of a spin state over others.

This finding uncovers a new facet of metal-porphyrin chemistry and, to some extent, ways to control it.
Our insights on the ferrous porphyrin may open new possibilities to manipulate 
electron delocalization, for instance via chemical functionalization of the macrocycle and have control over spin.
The role of peripheral functional groups in functionalized metal-porphyrin can be related to the proposed mechanism
and spin control can be achieved by \textit{ad hoc} groups that enhance or hinder electron delocalization.
Although the investigation focused exclusively on the ferrous porphyrin, our manuscript creates a paradigm that could extend to 
other transition metal porphyrins and therefore have a broader impact.
We predict that the same delocalization mechanism is responsible for the ability of
metal-porphyrins to undergo reduction and oxidation with ease in redox processes of living systems.
To date the nature of the oxidized or reduced species is poorly understood,
and so is the character of the acceptor orbital in reduction reactions.
According to the proposed mechanism, delocalization of the additional electrons (in the reduced form) is expected.
This delocalization of the extra charge reduces on-site repulsion making the event favorable.
\newpage

\section{Acknowledgments}
\label{Acknowledgments}

Roland Lindh and Ignacio F{\'e}rnandez Galvan are acknowledged for their assistence
in the evaluation of AO integrals within the Molcas chemistry package, via the Cholesky decomposition procedure.
Professors Marcel Swart and Kristine Pierloot are acknowledged for the inspiring discussions
on the relevance of accurate computational tools for spin chemistry investigations.
The Max-Planck Society and the COST Action CM1305 (ECOSTBio) are acknowledged for their financial support.

\section{Author contributions} \label{Contributions}

The authors have contributed equally in the investigation, conceiving ideas, developing the required theory and algorithm and planning and performing calculations.
Both authors have discussed the results and contributed to the final manuscript.

\section{Competing Financial Interests statement}
The authors declare no competing financial interest.

\section{Correspondence}
Correspondence and requests for materials should be addressed to \texttt{g.limanni@fkf.mpg.de} and \texttt{a.alavi@fkf.mpg.de}.

\section*{Supplementary materials}
The Supplementary Material contains details on:
Model System. Basis Set and electron repulsion integrals. Complete Active Space choices.
uantum Monte Carlo setup. Past and Present proposed active spaces.
Tables S1: Active Orbitals in D2h point group. Energy Splittings with VDZP basis set.
Figure S1: VDZP energy gaps. Listing 1: Cartesian Coordinates.
Figure S2: Natura orbitals of the quintet spin state. Mixing of d and $\pi$ orbitals.
Correlating d' orbitals. Wave function analysis. Listing 2: Molcas input example.
Listing 3: QMC input example. References cited within the Supporting Material.

%

\providecommand{\latin}[1]{#1}
\providecommand*\mcitethebibliography{\thebibliography}
\csname @ifundefined\endcsname{endmcitethebibliography}
  {\let\endmcitethebibliography\endthebibliography}{}

\newpage

\externaldocument[MA-]{main}
\newcommand{\beginsupplement}{%
        \setcounter{table}{0}
        \renewcommand{\thetable}{S\arabic{table}}%
        \setcounter{figure}{0}
        \renewcommand{\thefigure}{S\arabic{figure}}%
     }

\tikzset{
  elmove/.style={-{Stealth[#1]},shorten >=3pt,shorten <=2pt}
}

\lstset{
        commentstyle=\color{commentgreen},
        tabsize=4,
        basicstyle=\footnotesize\ttfamily,
        belowskip=-3pt,
        lineskip=-5pt    
}

\section{Supplementary Material.}

\beginsupplement
{\bf This PDF file includes:}
\begin{itemize}
\item Model System.
\item Basis Set and electron repulsion integrals.
\item Details on the Complete Active Space choice.
\item Details on the Quantum Monte Carlo setup.
\item Past and Present proposed active spaces.
\item Tables S1: Active Orbitals in D2h point group.
\item Energy Splittings with VDZP basis set.
\item Figure S1: VDZP energy gaps.
\item Listing 1: Cartesian Coordinates.
\item Figure S2: Natura orbitals of the quintet spin state.
\item Mixing of d and $\pi$ orbitals.
\item Correlating d' orbitals.
\item Wave function analysis.
\item Listing 2: Molcas input example.
\item Listing 3: QMC input example.
\item References cited within the Supporting Material 
\end{itemize}
\paragraph*{Model System.}
Our model system for the Fe(II)-porphyrin was derived from Pierloot's study\cite{Pierloot2017}.
The $\beta$-carbon atoms were removed and bonds saturated with hydrogen atoms (coordinates reported in Listing~1).
This simplification helped us to keep the calculations simple without removing the most important features of the system.
The Fe--N bonds were kept at a bond length of 1.989~$\AA$. The molecule was kept planar with $D_{4h}$ point group symmetry.
Aromaticity was preserved, with ring current and complete electron delocalization in the ``inner-cross'',
an 18 $\pi$ electrons system and 16 carbon atoms.
This simplification does not introduce bias towards the understanding of correlation effects in metallo-porphyrins.

The molecule was placed on the $xy$ plane, with the N atoms in between the $x$ and $y$ axes.
The D$_{2h}$ point group was used for all calculations, such that the $\pi-\pi^*$ orbital system belongs to
the b$_{1u}$, b$_{2g}$, b$_{3g}$ and a$_u$ irreducible representations. 
Orbitals $d_{z^2}$ and $d_{x^{2}-y^{2}}$ belong to the $a_g$ representation and,
the $d_{xy}$, $d_{xz}$ and $d_{yz}$ to the $b_{1g}$, $b_{2g}$ and $b_{3g}$ respectively.
Orbitals $d_{xz}$ and $d_{yz}$ and some of the $\pi-\pi^*$ orbitals belong to the same
irreducible representations ($b_{2g}$ and $b_{3g}$)
and, their overlap plays a major role in stabilizing the triplet spin state.

\paragraph*{Basis set and electron repulsion integrals.}
Generally-contracted atomic natural orbitals
(ANO-RCC) basis sets\cite{Widmark1990,Roos2005a} have been employed, obtained from the Fe(21s15p10d6f4g2h), C,N(14s9p4d3f2g), H(8s4p3d1f) primitive functions.
Two contraction schemes have been employed.
In one case the primitive functions have been contracted to Fe(5S4P2D1F), C,N(3S2P1D), H(2S1P),
giving a basis set of split-valence double-$\zeta$ plus polarization quality (VDZP).
In another case the primitive functions have been contracted to Fe(6s5p3d2f1g), C,N(4s3p2d1f), H(3s2p1d), 
giving a basis set of split-valence triple-$\zeta$ plus polarization quality (VTZP).
This second basis set choice led to a total of 707 basis functions.
Scalar relativistic effects were introduced via second order Douglas-Kroll-Hess integral correction.
The evaluation of the electron repulsion integrals has been greatly simplified by means of 
the resolution-of-identity Cholesky decomposition techniques 
as implemented in the Molcas software package\cite{JCC:JCC24221}, with a decomposition threshold of $10^{-4}$ a.u.
\cite{Aquilante2007a,Aquilante2007b,Aquilante2008a,Aquilante2009a,Aquilante2009b}.

\paragraph*{Details on the Complete Active Space choice.}
	CASSCF is probably the simplest and more natural method to tackle multi-configurational systems in chemistry
        \cite{Roos1980a,Roos1980b,Siegbahn1980,Siegbahn1981,Roos2007}.
        The core concept of CASSCF is the active space, a list of ``critical'' orbitals with their electrons for which a 
	complete many-body expansion is generated and orbitals are variationally optimized under 
	the field generated by the multiconfigurational wave function, removing any bias related
	to the choice of the trial orbitals. 
	Three main weaknesses of the CASSCF method need to be highlighted.
	(a) The larger the active space the exponentially larger the Configuration Interaction (CI) expansion, 
	limiting the active space to at most 18 active electrons and 18 active orbitals.
	(b) The active space represents a non-numerical parameter that introduces a certain level of arbitrariness and,
	active spaces that are smaller than the necessary might return wrong energetics 
	even upon perturbative correction. (c) Correlation outside the active space is completely 
	neglected at CASSCF level and post-CASSCF methods must be utilized for quantitative accuracy.

	Many methods have been proposed to reduce the exponential scaling of CAS wave functions. It is important to mention
	special forms of truncated CI expansions that can be obtained via the Restricted Active Space (RAS)\cite{Olsen1988,malmqvist1990,Celani2000} and
	the Generalized Active Space (GAS) approaches\cite{Ma2011,vogiatzis2015,Odoh2016}. 
        It is also important to mention the Density Matrix Renormalization Group (DMRG) approach
        \cite{white1992,Chan2002,White2005,Schollwock2005,Marti2008,chankallaygauss2004,chansharma2011,yanaikurashigeshoshchan2009}, 
        the variational two-electron reduced density matrices approach\cite{DePrince2016b,DePrince2016,mazziotti2012,Mazziotti2011,Mazziotti2008,mazziotti2007,mazziotti2006}
        and, the most recent Stochastic-CASSCF method\cite{Thomas2015,limanni2016}. The latter is the method of choice for this report\cite{limanni2016}.
	These methods partially circumvent the exponential scaling problem and enable the investigation of larger active spaces.
        Using massively parallelized architectures conventional CAS(20,20) calculations have recently been made possible\cite{vogiatzis2017}.
        When a sufficiently large active space is employed the bias due to the choice of the active space is to a great 
	extent lifted.
	In order to recover dynamic correlation outside the active space, perturbation theory to the second order (such as CASPT2
        \cite{andersson1990b,andersson1992b,andersson1993,finley1998,Ghigo2004,Pierloot2003,Pierloot2008,Pierloot2006,pulay2011,Vancoillie2010,Forsberg1997a},
        NEVPT2\cite{angeli2006-2,angeli2006-1,angeli2004,angeli2001,angeli2002,angeli2001-1,angeli2007}) and 
	multi-reference configuration interaction (MRCI) using CASSCF wave functions as reference,
	have been employed with great success in a wide range of chemical systems.
	RASPT2\cite{Sauri2011,Vancoillie2011,malmqvist2008} and GASPT2\cite{Ma2016} variants are also available.
	To date these methods represent the practical standard for transition metal chemistry.
	However, they become prohibitively expensive when coupled to reference wave functions built from large active spaces, 
	requiring in many cases further approximations, as discussed in great details in the literature\cite{Kurashige2011,Kurashige2014,Yanai2015,Phung2016}.
	Additionally, a second order approximation will not account for higher-order correlation processes and 
	orbital relaxation (\emph{vide intra}).
	The Multi-Configuration Pair-Density Functional Theory (MCPDFT) method has
	been proposed as a cheap alternative to CASPT2\cite{limanni2014,carlson2015a,Odoh2016,gagliardi2017,Hoyer2016,Wilbraham2017,Ghosh2015,Hoyer2015,Ghosh2017}.
	Its computational cost is nearly independent of the size of the underlying
	reference wave function and therefore nearly insensitive to the size of the active space.

For the CASSCF calculations several active spaces have been chosen:
(a) The CAS(6,5) is the smallest active space that includes solely the six valence electrons of the
metal centre and its five 3d orbitals.
(b) In the CAS(8,6) a doubly occupied $\sigma$ orbital is added, mostly localized on the N atoms
and pointing to the direction of the $3d_{xy}$ orbital of the iron atom.
(c) The CAS(8,12) adds five empty correlating $d'$ and the Fe 4s orbitals into the active space.
The Fe 4s has been added as it could compete with the $d'$ orbitals in accounting for correlation effects.
(d) In the CAS(14,16), the role of the frontier $\pi$ orbitals was probed. 
The two highest bonding $\pi$ orbitals and the two lowest anti-bonding $\pi*$ orbitals
have been added to Pierloot CAS(8,11) active space together with their four electrons.
We also added the doubly occupied 3s orbital on the Fe centre to investigate
its role in the electron correlation landscape.
(e) The much larger active space, CAS(32,34), consists of the 10 Fe ($3d$,$d'$)
orbitals and their 6 electrons, the entire $\pi$ system (18 electrons and 16 orbitals),
the four orbitals of the Fe (4s4p) shell and four doubly occupied N ($2p_{x}$,$2p_{y}$)
symmetrically combined ``radial'' orbitals pointing at the metal centre.
The four remaining N ($2p_{x}$,$2p_{y}$) orbitals, symmetrically combined
to form ``tangential'' orbitals, were not included in the active space.
The orbitals correlated in the CAS(32,34) are different than
the ones used in our previous work\cite{limanni2016}.
For the CAS(32,29), only valence orbitals on the macrocycle and the metal centre were included.
In the present work double-shell orbitals, ``radial'' N 2p orbitals and, the Fe ($4s$,$4p$) orbitals have
been explicitely correlated.
The enlarged active space shows the \textit{breathing mechanism} and provides correct energy ordering of the spin states.

For the small active spaces (cases (a) to (d) above) the CASPT2 method has been used to recover 
dynamic correlation outside the active space.
At CASPT2 level, core orbitals ($1s$ on C and N atoms and $1s$, $2s$, $2p$ 
orbitals on Fe atom) were kept frozen.
The standard IPEA zeroth order Hamiltonian has been utilized with the default IPEA denominator shift of 0.25~a.u.
No method for dynamic correlation has been coupled to the Stochastic-CASSCF wave functions.
	Calculations at CASPT2 level show that predictions are largely 
	affected by the choice of the under-lying active space with the high 
	spin state still over-stabilized over the triplet spin state.
For comparison with the Stochastic-CAS(32,34) results, also RASSCF calculations have been performed,
in which 32 electrons and 34 active orbitals have been partially correlated.
Following Pierloot's approach, in our RAS(32,34) twelve orbitals were put in the RAS1 space,
including three $\sigma$ (N 2p) orbitals and, nine $\pi$ orbitals, six orbitals in RAS2,
including one bonding $\sigma$ orbital and the five 3d orbitals and, sixteen orbitals in RAS3,
including five double-shell correlating orbitals and the $\pi^{*}$ orbitals of the macrocycle.
Only single and double excitations out of RAS1 and into RAS3 have been allowed.

\paragraph*{Details on the Quantum Monte Carlo setup.}
For the FCIQMC dynamics the \emph{initiator} formulation of the method has been used \cite{Alavi2010,Alavi2011} with
a threshold value of $n_a=3.0$ together with the semi-stochastic method
\cite{umrigar2012,bluntsmart2015} using a deterministic subspace consisting
of $|{\cal D}|=10000$ most populated determinants. 
In the initiator approximation, walkers populating determinants with largest weight are promoted to be ``initiators''.
Only initiators are able to spawn walkers on empty determinants. Non-initiators are allowed to spawn only 
on determinants that are already occupied.
The calculations were run in replica mode\cite{Alavi2014} in order to sample the one- and 
two-body reduced density matrices necessary to the orbital rotation step.
CASSCF natural orbitals from smaller active spaces were used as the
starting orbitals for CASSCF optimizations with larger active spaces.
$5\times 10^6$ walkers were employed for the initial dynamics and the first five CASSCF iterations.
The number of walkers was gradually increased to $5\times 10^8$. 
CASSCF convergence was reached at this walker population for all states here investigated.
The approach of increasing the walker population in steps follows from our initial findings, 
already discussed in our previous paper\cite{limanni2016}. A small walker distribution is able to generate
a convenient averaged field to allow for an effective orbital optimization step
at the early stages of the CASSCF procedure.
This procedure guarantees fast orbital rotations and a limited number of 
CASSCF iterations at the high-population regime when sub-milliHartree accuracy is required.
After CASSCF convergence was reached, more refined solutions were obtained by
increasing the target number of walkers to $1\times 10^9$.
This procedure is standard to reduce the initiator error on the stochastic sampling of the wave function and,
was used to confirm that no bias on the spin splitting was introduced due to undersampling of the determinantal space.
The time-step $\Delta \tau$ was found via an automatic search
procedure \cite{boothsmartalavi2014} for each simulation, and took typical
values in the range $5-10\times 10^{-4}$ a.u.
A typical FCIQMC simulation, took $\sim24$ hours for each CASSCF iteration on 640 cores.
Orbital rotations were performed using the Super-CI method with a quasi-Newton update.
The entire CASSCF procedure converged in 10-15 iterations for the states here investigated.
All calculations have been performed using the OpenMolcas chemistry software package\cite{JCC:JCC24221}.

\paragraph*{Past and present proposed active spaces.}
        The smallest active space for this system would be a CAS(6,5) 
	including only the six valence electrons in the five valence 3d orbitals. Pierloot pointed out that one additional 
	$\sigma$ Fe--N bonding orbital, and its two electrons, must be included in the active space, leading to a CAS(8,6).
	She also found that it is crucial to include five double-shell correlating orbitals in the active space
	when a PT2 treatment is used for dynamic correlation. These orbitals account for a quite strong radial
 	electron correlation and, when included into the active space, lead to a CAS(8,11).
	Recently Pierloot and co-workers have investigated the role of the (3s3p) and (4s4p) shells to give
	a CAS(16,15) and a CAS(16,19), respectively. For the latter being prohibitively large, the RASSCF/RASPT2
	or the DMRG/PT2 methodologies have been used\cite{Pierloot2017}.
	Pierloot has also considered the role of the $\pi$-system of the macrocycle\cite{Vancoillie2011}, 
	by means of the RASSCF/RASPT2 approach. A RAS(34,2,2;13,6,16) has been chosen containing a total of 34 electrons 
	and 35 orbitals. RAS1 contains 13 active orbitals, doubly occupied in the reference determinant, RAS2 contains only 6 
        active orbitals and RAS3 the remaining 16 that are empty in the reference determinant.
	Only single and double excitations were allowed from RAS1 and to RAS3.

	Among the successful wave function methods able to predict a triplet ground state, it is important to mention
	Rado{\'n}'s CCSD(T) calculations, that have shown a triplet ground state only
	upon extrapolation to the complete basis set limit\cite{Radon2014}, a DMRG-CI study\cite{Olivares2015} and,
        the recent Heat-bath Configuration Interaction Self-Consistent Field (HCISCF) by Sharma\cite{Sharma2017}.
	In the last two cases a CAS(44,44) was chosen.

The CAS(32,34) used in the present report differs from the one used in the past\cite{Olivares2015,Sharma2017}.
The CAS(44,44) reported in the literature includes $4p_x$ and $4p_y$ orbitals of the metal centre but,
does not include the $4s$ or the $4p_{z}$ orbitals. Therefore, the (4s,4p) shell is somehow incomplete. 
Olivares-Amaya and Sharma included the entire list of eight MOs resulting from a symmetry adapted combination
of the $2p_x$ and $2p_y$ orbitals on the N atoms. In the present work only four have been included, leaving the tangential ones in the inactive space.
No orbitals have been frozen or deleted at the CASSCF level of theory. 

	In the paper introducing the Stochastic-CASSCF method, we performed calculations on a model system of Fe(II)-porphyrin\cite{limanni2016}.
	Our model system had a Fe--N bond length of 2.05~\AA, which is closer to the one predicted for the quintet $^{5}A_{1g}$ state.
	An active space of 32 electrons in 29 orbitals was chosen, including the 24 orbitals of the aromatic $\pi$-system on
	the macrocycle with their 26 electrons and the five 3d valence orbitals of the metal centre with their six electrons.
	The Stochastic-CASSCF(32,29) led to a quintet ground state ($^{5}A_{1g})$ with the triplet at ~14~kcal/mol.
	Neither the double-shell orbitals, nor the bonding Fe--N $\sigma$ orbital were added to the active space.
	Inspection of the CAS(32,29) wave function revealed that the dominant configuration
	$(3d_{z^{2}})^{2} (3d_{x^{2}-y^{2}})^{1}(3d_{yz})^{1}(3d_{xz})^{1}(3d_{xy})^{1}$, contributed for 37.3\%.
	The other relevant configurations (with weight $>5\%$) were charge-transfer
	configurations in which electrons from the metal centre are excited into
	the $\pi$ orbitals.  Older investigations concluded that the iron 3d orbitals 
	and the porphyrin $\pi$ system were well separated, 
	whereas our findings showed the opposite, with the Stochastic-CASSCF wave 
	function showing a close interaction between the aromatic ring and the metal centre 
	via non-negligible charge-transfer configurations.

The distribution of orbitals for the CAS(32,34) proposed here in the eight irreducible representations 
of the $D_{2h}$ point group is given in Table~\ref{Tab:Orbitals}.
\begin{table}[!ht]
    \centering
    \caption{%
        Distribution of molecular orbitals among inactive, active and secondary
        spaces for each irreducible representation. 
    }
    \setlength{\tabcolsep}{0.1em}
    \begin{tabular*}{0.8\textwidth}{@{\extracolsep{\fill}}ccccccccc}
                          &  Ag & B3u & B2u & B1g & B1u & B2g & B3g & Au \\
        \hline            
        Inactive          &  16 &  13 &  13 &  9  &  2  &  0  &  0  &  0 \\
        Active            &   6 &   2 &   2 &  3  &  6  &  6  &  6  &  3 \\
        Secondary VTZP    & 113 & 104 & 104 & 93  & 59  & 51  & 51  & 45 \\
        \hline            
    \end{tabular*}          
    \label{Tab:Orbitals}
\end{table}

\begin{figure}[H]
         \centering%
         \includegraphics[width=16cm]{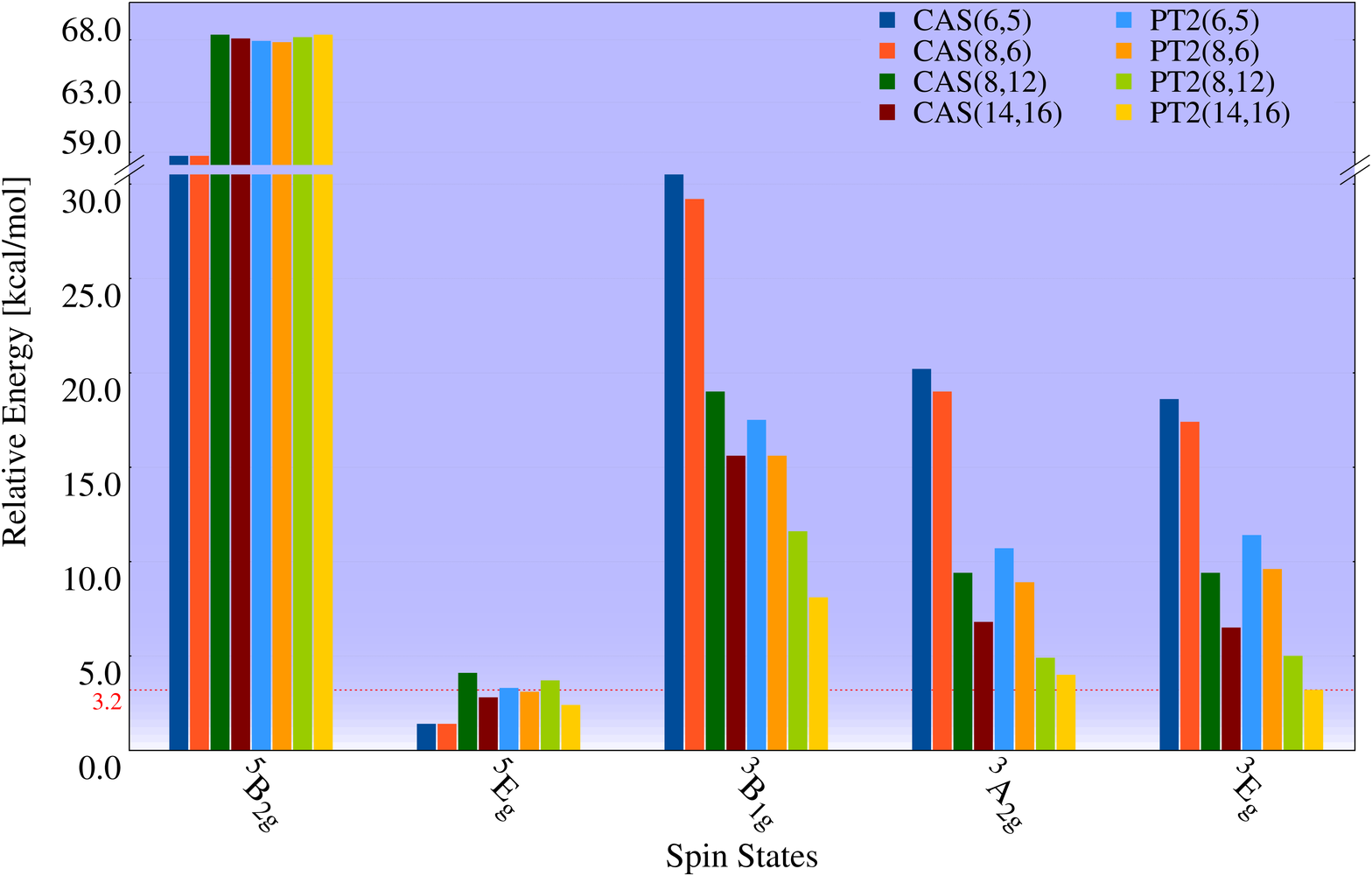}
	 \caption{Energy splittings relative to the $^{5}A_{1g}$ state in VDZP basis set. The red dashed line marks the lowest triplet-quintet energy gap obtained.%
         }%
         \label{Fig:VDZP}
     \end{figure}

\paragraph*{Energy Splittings with VDZP basis set.}
Energy splitting estimates at the various level of theory and VDZP basis set are summarized in Figure~\ref{Fig:VDZP}.
A comparison with Figure~2 of the main text (where a larger basis set of TZVP quality is used) shows the effect of the basis set on the spin ordering.
\newpage

\begin{lstlisting}[frame=single,label=coord,caption = Cartesian coordinates for the Fe(II)-porphyrin model system (Angstrom)]
   29

Fe     0.000000     0.000000     0.000000
 N     1.406727     1.406727     0.000000 
 N    -1.406727     1.406727     0.000000
 N     1.406727    -1.406727     0.000000
 N    -1.406727    -1.406727     0.000000
 C    -0.000000     3.400142     0.000000
 C    -0.000000    -3.400142     0.000000
 C     3.400142    -0.000000     0.000000
 C    -3.400142    -0.000000     0.000000
 C     1.222770     2.760387     0.000000
 C    -1.222770     2.760387     0.000000
 C     1.222770    -2.760387     0.000000
 C    -1.222770    -2.760387     0.000000
 C     2.760387     1.222770     0.000000
 C    -2.760387     1.222770     0.000000
 C     2.760387    -1.222770     0.000000
 C    -2.760387    -1.222770     0.000000
 H     0.000000     4.482672     0.000000
 H     0.000000    -4.482672     0.000000
 H     4.482672     0.000000     0.000000
 H    -4.482672     0.000000     0.000000
 H     2.181081     3.277651     0.000000
 H    -2.181081     3.277651     0.000000
 H     2.181081    -3.277651     0.000000
 H    -2.181081    -3.277651     0.000000
 H     3.277651     2.181081     0.000000
 H    -3.277651     2.181081     0.000000
 H     3.277651    -2.181081     0.000000
 H    -3.277651    -2.181081     0.000000
\end{lstlisting}

    \begin{figure} []%
        \captionsetup[subfigure]{labelformat=empty,farskip=-25pt,captionskip=3pt}
        \centering%
         \includegraphics[width=14cm]{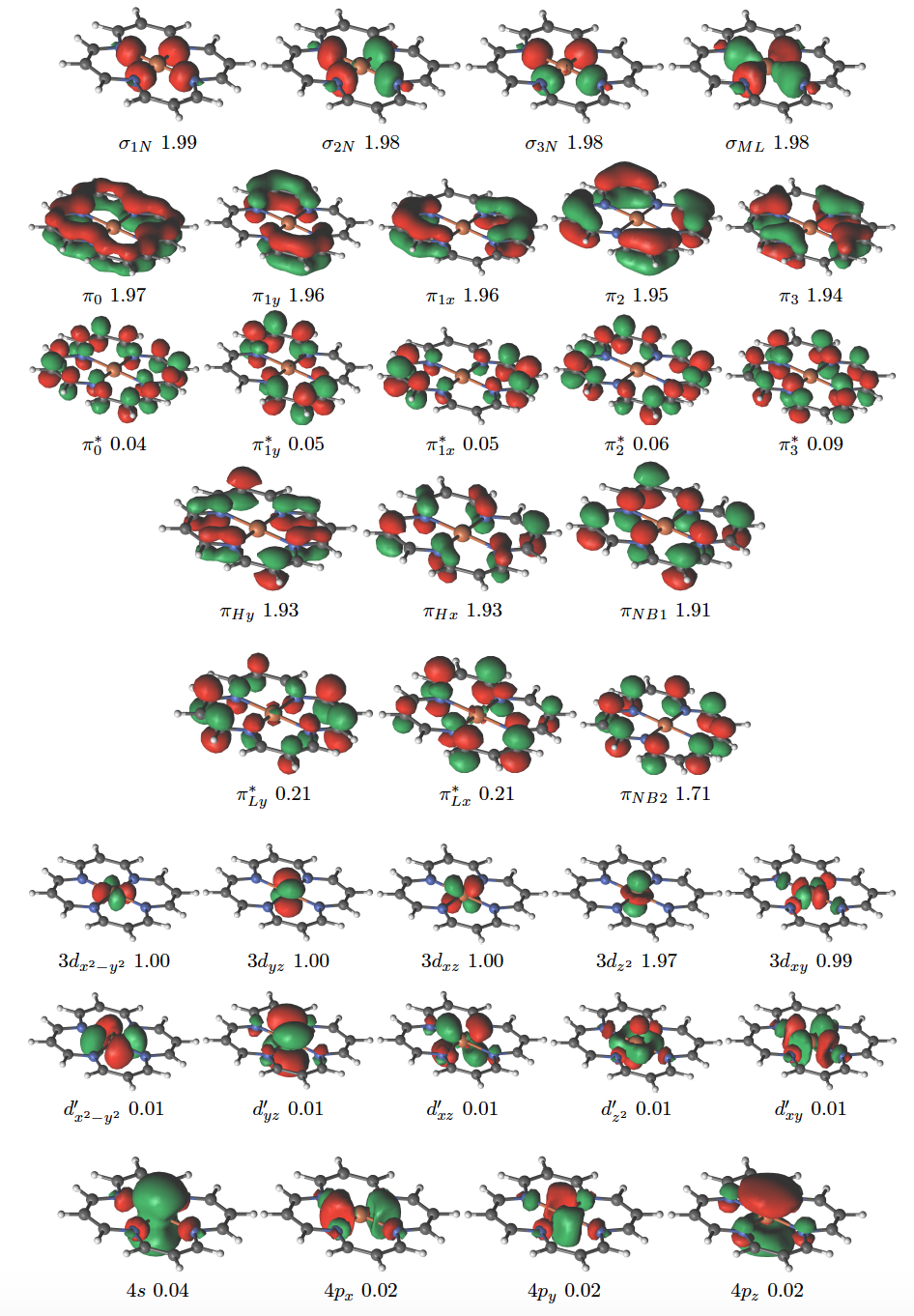}
        \caption{%
            Stochastic-CAS(32,34) active natural orbitals for the $^{5}A_{1g}$ state of the Fe(II)-porphyrin model system
	    and their occupation numbers.
        }
        \label{SI:5A1g_NatOrb3234}
    \end{figure}
\newpage

\paragraph*{On the mixing of metal-d and $\pi$ orbitals.}
In order to understand the importance of our results on the mixing of the metal and ligand natural orbitals for the triplet spin state 
and not for the quintet spin state, we would like to comment on some properties of the CASSCF approach.
CASSCF is invariant to orbital rotations within the active space and thus rotations within the active orbitals
do not alter the energy gap between the states. Therefore any sort of unitary transformation acting on
the active orbitals would not change the physics of the states. Here we use natural orbitals.
Natural orbitals follow a strict recipe -- they diagonalize the reduced one-body density matrix --
that leads to a uniquely defined set of rotated CASSCF active orbitals. 
When occupation number degeneracies arise any rotations within orbitals with same occupancy is possible
and in this case natural orbitals can be presented in various ways.
This special case does not arise in the current investigation.

In computing the natural orbitals one may think of a two step procedure.
In a first step, active orbitals are localized in a way that
each of them is either sitting on the metal centre or on the macrocycle.
This step is conceptually important to be able to clearly state in which
part of the molecule one electron (or pair of electrons) is located;
however, in practice, it is not needed as the shape of the natural orbitals
will not be affected by the starting orbitals used to build the one-body density matrix.
In a second step, starting from these localized orbitals,
the one-body density matrix is built and diagonalized.
Diagonalization of the one-body density matrix is the responsible step that leads (or not) to the mixing of the localized orbitals.
If non negligible off-diagonal elements of the one-body density matrix exist, they will have an effect on the mixing.
Non-negligible off-diagonal elements can exist only for correlated systems.
This is exactly what we observe in the system under investigation. $\pi$ orbitals and out-of-plane 3d orbitals are correlated via charge-transfer excitations,
they will lead to non-negligible off-diagonal elements in the one-body density matrix and thus to the mixing of the metal centre and the ligand orbitals.

\paragraph*{Notes on the correlating d' orbitals.}
In our active space we added a set of five correlating d' orbitals. These are known as double-shell $d'$ orbitals.
They have a nodal structure that resemble the 4d physical orbitals.
Nonetheless, their radial distribution is closer to the 3d orbitals and their energy is higher than the 4d orbitals.
They could be defined as contracted 4d orbitals. This feature is very general and not specific to the present compound.
CASSCF forces them to be closer to the 3d orbitals to maximize their overlap and the radial correlation of the 3d electrons.
They are responsible for the \textit{breathing effect}, according to which valence electrons of the correlated method 
are more expanded than the equivalent electron in a non correlated (or less correlated) approach.
This breathing mechanism leads to the reduction of the on-site electron repulsion of the doubly occupied 3d orbitals of the metal centre.

\paragraph*{Wave function analysis.} The $^{3}E_{g}$ state is dominated by the
$(\sigma_{N})^{6}(\pi)^{18}(\sigma_{ML})^{2}(3d_{x^{2}-y^{2}})^{2}$ $(3d_{xz},3d_{yz})^{3}(3d_{z^{2}})^{1}$
configuration with a weight of 59\%.
The second two most relevant configurations (contributing for 3\% and 2\% respectively)
consist of double excitations from the $\pi_{NB2}$ to the $\pi^{*}_{Lx}$ and the $\pi^{*}_{Ly}$ respectively.
These two configurations are the main responsible for populating the $\pi^{*}_{Lx}$ and the $\pi^{*}_{Ly}$ natural orbitals.
The same configurations contribute for less than 0.5\% in the equivalent RAS wave function.
$\pi \rightarrow \pi^{*}$ excitations contribute for $\sim 1\%$ to the wave function.
$\sigma_{ML}\rightarrow 3d$ and $3d\rightarrow 3d$ excitations also appear in the leading determinant of the triplet spin state.
    The $^{5}A_{1g}$ state is dominated by the $(\sigma_{N})^{6}(\pi)^{18}(\sigma_{ML})^{2}(3d_{z^{2}})^{2}(3d_{x^{2}-y^{2}})^{1}(3d_{xz})^{1}$ $(3d_{yz})^{1}(3d_{xy})^{1}$ configuration with a weight of 61\%.
    Other relevant configurations are the double excitations from the $\pi_{NB2}$ and $\pi_{NB1}$ to the $\pi^{*}_{Lx}$ and the $\pi^{*}_{Ly}$ respectively contributing for 3\% each to the wave function.
    The $3d_{z^{2}}\rightarrow 3d_{x^{2}-y^{2}}$ and $3d_{z^{2}}\rightarrow 4s$ excitations contribute for $\sim$~0.4\% each.
    Lower contributions to the wave function come from other $\pi \rightarrow \pi^{*}$ (ring delocalization) and $3d \rightarrow d'$ excitations (radial correlation).
\newpage

\begin{lstlisting}[frame=single,label=MolcasInp,caption = A possible FCIQMC-CASSCF input setup for the porphyrin molecule within OpenMolcas]
&GATEWAY
RICD
Coord
3A2g.xyz
basis
ANO-RCC-VTZP
group
full

&SEWARD

&RASSCF
NECI
EXNE
NWAL
50000000
DEFD
1 2 3 4 5 13 14 17 18 21 22 27 28 29 30 31 
32 39 40 41 42 43 51 52 53 54 55 56 63 64 65 66
CYCLe
200000
RSPCutoff
0.3
TIMENeci
2000
RDMStart
50000
RDMPick
1000
THRS
1.0e-4 1.0e-1 5.0e-4
spin
3
Symmetry
6
nactel
32 0 0
inactive 
 16 13 13 9 2 0 0 0 
ras2
  6  2  2 3 6 6 6 3
deleted
  0  0  0 0 0 0 0 0 
\end{lstlisting}

\begin{lstlisting}[frame=single,label=QMCinp,caption = Relevant input keywords used by the FCIQMC program NECI]
Title
System read
electrons  32
nonuniformrandexcits 4ind-weighted-2
nobrillouintheorem
spin-restrict    2
freeformat
endsys
 
calc
definedet    1    2    3    4    5   13   14   17   18   21   22   27   28   29   30   31   32   39   40   41   42   43   51   52   53   54   55   56   63   64   65   66

methods
method vertex fcimc
endmethods
 
totalwalkers    1000000000
diagshift 0.00
shiftdamp 0.02
nmcyc     200000
stepsshift 10
proje-changeref 1.2
truncinitiator
addtoinitiator 3
allrealcoeff
 realspawncutoff 0.30
jump-shift
tau 0.001 search
max-tau 0.02
maxwalkerbloom 1
memoryfacspawn 10.0
memoryfacpart 5.0
time 1400
startsinglepart 100
readpops
walkcontgrow
semi-stochastic
pops-core 10000
trial-wavefunction
pops-trial 500
rdmsamplingiters 10000
endcalc
 
logging
PRINT-SPIN-RESOLVED-RDMS
(READRDMS
HDF5-POPS 
Highlypopwrite 200
(binarypops
printonerdm
diagflyonerdm
calcrdmonfly          3     5000      500
endlog
end
\end{lstlisting}

%

\providecommand{\latin}[1]{#1}
\providecommand*\mcitethebibliography{\thebibliography}
\csname @ifundefined\endcsname{endmcitethebibliography}
  {\let\endmcitethebibliography\endthebibliography}{}

\end{document}